\documentclass[english,aip,jmp,amssymb,amsmath,reprint]{revtex4-1}
\usepackage[T1]{fontenc}
\usepackage[utf8]{inputenc}
\usepackage{amsmath}
\usepackage{amssymb}
\usepackage{mathrsfs}
\usepackage{graphicx}
\usepackage{esint}
\usepackage{babel}



\def \beq {\begin{equation}}
\def \edq {\end{equation}}
\def \bes {\begin{subequations}}
\def \eds {\end{subequations}}
\def \beqn {\begin{equation*}}
\def \edqn {\end{equation*}}

\def \dag {\dagger}

\def \up {\uparrow}
\def \down {\downarrow}

\def \sm {\sigma}

\def \veps {\varepsilon}

\def \calh {{\cal{H}}}

\def \calt {{\cal{T}}}

\def \cala {{\cal{A}}}

\def \scrg {{\mathscr{G}}}

\def \tGamma {\widetilde{\Gamma}}

\def \hc {\text{H.c.}}

\begin{document}

\title{Proposal for a local heating driven spin current generator}

\author{Sun-Yong Hwang}

\affiliation{Department of Physics, Pohang University of Science and Technology, Pohang 790-784, Korea}

\affiliation{Institut de F\'isica Interdisciplin\`aria i Sistemes Complexos IFISC (UIB-CSIC), E-07122 Palma de Mallorca, Spain}

\author{Jong Soo Lim}

\affiliation{Institut de F\'isica Interdisciplin\`aria i Sistemes Complexos IFISC (UIB-CSIC), E-07122 Palma de Mallorca, Spain}

\author{Rosa L\'opez}

\affiliation{Institut de F\'isica Interdisciplin\`aria i Sistemes Complexos IFISC (UIB-CSIC), E-07122 Palma de Mallorca, Spain}

\affiliation{Departament de F\'isica, Universitat de les Illes Balears, E-07122 Palma de Mallorca, Spain}

\author{Minchul Lee}

\affiliation{Department of Applied Physics, College of Applied Science, Kyung Hee University, Yongin 446-701, Korea}

\author{David S\'anchez}

\affiliation{Institut de F\'isica Interdisciplin\`aria i Sistemes Complexos IFISC (UIB-CSIC), E-07122 Palma de Mallorca, Spain}

\affiliation{Departament de F\'isica, Universitat de les Illes Balears, E-07122 Palma de Mallorca, Spain}

\begin{abstract}
We propose a two-terminal spin-orbit interferometer with a hot molecule inserted
in one of its arms to generate pure spin currents. Local heating is achieved
by coupling the vibrational modes of the molecule to a third (phononic) reservoir.
We show that this spin calorimetric effect is due to the combined influence of
spin-dependent wave interference and inelastic scattering.
Remarkably, the device converts heat flow into spin-polarized current
even without applying any voltage or temperature difference to the electronic
terminals. 
\end{abstract}

\maketitle

Recent experimental demonstrations of spin-polarized currents
using thermal gradients only\cite{uch08,sla10} has fueled the interest
in finding synergies between thermoelectricity and spintronics.
Thus, the field of spin caloritronics\cite{bau10} seeks 
new functionalities that exploit the coupling of
charge, spin and energy degrees of freedom in nanostructures.
Here we propose a molecule-based spin caloritronic device
that extracts heat from a nearby phonon bath
and transforms it into a spin current that flows out into
coupled electronic reservoirs. Crucial to our setup
is the presence of tunable spin-orbit interactions
that causes traveling electrons to acquire a phase
which depends on its spin orientation within
an Aharonov-Bohm-type interferometer.

Recent works predict that heat current can be converted
into electric current in three-terminal thermoelectric
nanodevices either due to inelastic processes
at a molecular bridge\cite{ent10} or due
to Coulomb coupling in interacting quantum dots.~\cite{san11}
The generated charge current is determined by
the temperature difference between the 
third terminal kept at an elevated temperature
and the base temperature of the system.
Further investigations include
time-reversal symmetry breaking effects,\cite{ent12}
chaotic cavity heat engines,\cite{sot12}
and phonon-assisted instabilities.\cite{sim12}
The effect originates from a rectification
of temperature fluctuations in the coupled
unbiased conductor, similarly to the directed
motion induced in the drag effect.\cite{san10}
Importantly, unlike the pure electric case which requires
four current-carrying terminals, in these thermoelectric effects
it suffices to couple a third energy-supplying
subsystem (electronic or phononic).
\begin{figure}[!h]
\begin{center}
\includegraphics[width=9cm]{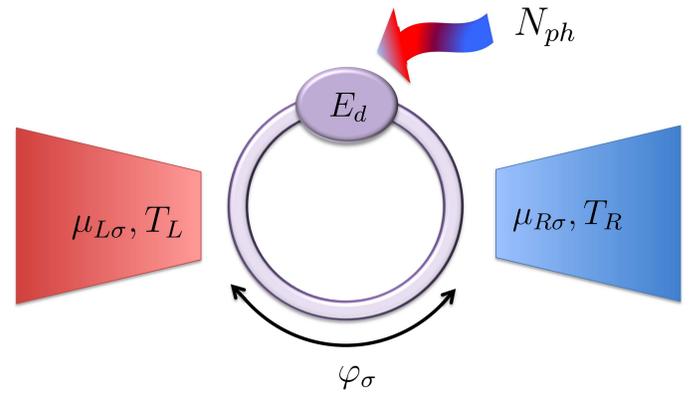}
\caption{Schematics of the setup. A molecule or quantum dot with a single level $E_d$
is embedded in an Aharonov-Bohm ring coupled to two
Fermi reservoirs ($L$ and $R$) with electrochemical potentials $\mu_{L(R)}$
and temperatures $T_{L(R)}$. A spin-dependent phase $\phi_\sigma$
originates from spin-orbit Rashba interactions, generally
leading to spin-dependent chemical potentials, $\mu^\sigma$. 
Our device is influenced by the presence of a bosonic bath described by the Bose-Einstein distribution function $N_{ph}$
with temperature $T_{P}$. }
\label{fig:scheme}
\end{center} 
\end{figure}

Our proposal considers an Aharonov-Bohm (AB) ring connected
to two normal electronic reservoirs
and a molecule embedded in one arm of the ring, see Fig.~1.
For definiteness, the molecule is assumed
to have a single level active for transport. The system is subjected to Rashba spin-orbit
interaction arising in inversion asymmetric potentials. This configuration
has been shown to give rise to local spin polarizations,\cite{sun06}
spin separation,\cite{chi08} and pure spin currents.\cite{lu07}
Furthermore, this is a useful setup to analyze the competition between
spin randomizing Rashba interactions and many-body spin singlets
arising from Kondo correlations.\cite{ver09,lim10}
These works require small voltage biases applied
to the Fermi reservoirs. In contrast, our device operates
in the entire absence of electric or thermal gradients applied
to the leads but is instead based on a hot phonon field locally
coupled to the embedded molecule.

The model Hamiltonian for our device, $\calh = \calh_C + \calh_M + \calh_T$, consists of three parts:
$\calh_C = \sum_{\ell=L/R,k,\sm} \veps_{k\sm} c_{\ell k\sm}^{\dag}c_{\ell k\sm}$ is the Hamiltonian of two normal left ($L$) and right ($R$) electronic reservoirs,
$\calh_M = \sum_{\sm} E_d d_{\sm}^{\dag}d_{\sm} + \hbar\omega_0 a^{\dag}a + \lambda(a+a^{\dag})\sum_{\sm} d_{\sm}^{\dag}d_{\sm}$ describes 
the embedded molecule with level $E_d$ coupled to a phonon bath with an excitation frequency $\omega_0$ and coupling strength $\lambda$, and $\calh_T = \sum_{\ell,k,\sm} \left(V_{\ell} c_{\ell k\sm}^{\dag}d_{\sm} + \hc\right) + \sum_{k,p,\sm} \left(We^{i\varphi_{\sm}}c_{Rp\sm}^{\dag}c_{Lk\sm} + \hc\right)$
accounts for electron tunneling between the molecule and the reservoirs.
Here, $c_{\ell k\sm} (c_{\ell k\sm}^{\dag})$  are fermionic annihilation (creation) operators
for electrons with wave vector $k$ and spin $\sigma$ in lead $\ell=L/R$
and $d_{\sm} (d_{\sm}^{\dag})$ represent electrons at the molecule site and
$a (a^{\dag})$ denotes the bosonic annihilation (creation) operator. 
$V_{\ell}$ is the probability amplitude for an electron transfer between the molecule and the electronic reservoirs
whereas $W$ describes direct tunneling between electronic reservoirs.
The effect of Rashba spin-orbit interaction is embodied in the phase factor\cite{sun06,chi08,lu07,ver09,lim10}
$\varphi_{\sm} = \sm\varphi$, where $\varphi= \alpha_R l$ ($\alpha_R$ is the Rashba strength and $l$ the size of the molecule)
and $\sigma=+$($-$) for spins up (down).

We consider thermoelectric effects in the linear response regime. Thus,
we expand the electrochemical potential and temperature in each electronic reservoir
around the equilibrium state defined with common chemical potential $\mu$ and base
temperature $T$. Then,
$\mu_{\ell\sm} = \mu + \Delta\mu_{\ell\sm}$ and $T_{L/R} = T \pm \Delta T/2$;
similarly, $T_P = T + \Delta T_P$ for the phonon bath. 
Using $\Delta\mu_{\sigma} = \mu_{L\sigma} - \mu_{R\sigma}$, we define the charge and spin voltage biases $e\Delta V = (\Delta\mu_{\up} + \Delta\mu_{\down})/2$ and $e\Delta V_s = (\Delta\mu_{\up} - \Delta\mu_{\down})/2$, respectively.\cite{swi09}
Hence, the spin-resolved current is
\begin{equation}\label{eq_Ism}
I_{\sm} = G(\varphi_{\sm}) \Delta V + \sm G(\varphi_{\sm}) \Delta V_s + L(\varphi_{\sm})\frac{\Delta T}{T} + X_{P}(\varphi_{\sm})\frac{\Delta T_P}{T}.
\end{equation}
The transport coefficient $G$ is the linear conductance in response to a charge or spin voltage shift.
$L$ is the thermoelectric response due to a temperature difference applied to the electronic leads.
Finally, $X_P$ describes phonon-assisted transport owing to possible temperature differences
between the bosonic bath and the system. We remark that local phonon heating effects have been investigated
in electronic molecular systems.\cite{fre04,gal07} 

Importantly, the transport coefficients in Eq.~\eqref{eq_Ism} have elastic and inelastic contributions: 
$G= G_{el} + G_P$ and $L = L_{el} + L_P$. Expressions for $G_{el}$ and $L_{el}$ can be found
in Ref.~\onlinecite{ent12} with the replacement $\varphi_{AB} \to \varphi_{\sm}$.
We here prefer to focus on the inelastic contribution originated from electron-phonon (e-ph) coupling
since it is precisely the combined effect of inelastic scattering and spin-orbit interaction
which establishes the operating principle of our proposed device. Within the nonequilibrium
Green function approach,\cite{haug} we find
$G_P(\varphi_{\sm}) = (2\lambda^2 e^2/h) \int d\veps B(\veps,\varphi_{\sm})M_0(\veps,\varphi_{\sm})$,
where $B(\veps,\varphi_{\sm}) = (N_{T}/k_BT) \tGamma(\veps_{+})\tGamma(\veps_{-})|\scrg_{\sm,\sm}^r(\veps_{+})\scrg_{\sm,\sm}^r(\veps_{-})|^2(1-f_0(\veps_{+}))f_0(\veps_{-})$ and
$M_0(\veps,\varphi_{\sm}) \equiv 1 - \cala(\veps_+)\cala(\veps_-) + \sqrt{\calt(\veps_+)\calt(\veps_-)}\sin^2(\varphi_{\sm})$
with $\veps_{\pm} = \veps \pm \hbar\omega_0/2$, the Fermi distribution function
$f_0(\veps) = (e^{(\veps-\mu)/k_BT}+1)^{-1}$, and the Bose occupation factor $N_T = (e^{\hbar\omega_0/k_BT}-1)^{-1}$,
both evaluated at equilibrium. 
Here, $\tGamma(\veps) = (\Gamma_L(\veps) + \Gamma_R(\veps))/(1+\xi(\veps))$ is the (renormalized) total resonance width, where
$\Gamma_{\ell}(\veps) = \pi\rho_{\ell}(\veps) V_{\ell}^2$ and $\xi(\veps) = \pi^2\rho_L(\veps)\rho_R(\veps) W^2$
is the direct tunneling coupling, with $\rho_{\ell}(\veps)$ the $l$-lead density of states.
The retarded Green's function in the absence of e-ph coupling is given by $\scrg_{\sm,\sm}^r(\veps) = (\veps-\veps_d + i\tGamma+ \tGamma\sqrt{\alpha\xi}\cos\varphi_{\sm})^{-1}$ 
with $\alpha(\veps) = 4\Gamma_L\Gamma_R/(\Gamma_L+\Gamma_R)^2$. We have defined the functions
$\cala(\veps) = [(\Gamma_L-\Gamma_R)/(\Gamma_L+\Gamma_R)][(1-\xi)/(1+\xi)]$
and $\calt(\veps) = 4\alpha\xi/(1+\xi)^2$. 
The thermopower inelastic term can be cast in the form 
$L_P(\varphi_{\sm}) = (2\lambda^2e/h)\int d\veps [(\veps-\mu)B(\veps,\varphi_{\sm})M_0(\veps,\varphi_{\sm})+(\hbar\omega_0/2)M_1(\veps,\varphi_{\sm})]$,
where $M_1(\veps,\varphi_{\sm}) \equiv \cala(\veps_+)\sqrt{\calt(\veps_-)} - \cala(\veps_-)\sqrt{\calt(\veps_+)}\sin\varphi_{\sm}$.
Finally, the phonon assisted current contribution is
$X_P(\varphi_{\sm}) = (2\lambda^2 e/h)\hbar\omega_0 \int d\veps B(\veps,\varphi_{\sm})M_2(\veps,\varphi_{\sm})$ with
$M_2(\veps,\varphi_{\sm}) \equiv \left[\cala(\veps_-) - \cala(\veps_+)\right] + \left[\sqrt{\calt(\veps_+)} - \sqrt{\calt(\veps_-)}\right]\sin\varphi_{\sm}$.

In the wide-band limit, $\rho_{\ell}$ is energy independent.
Then, $\calt(\veps)$ and $\cala(\veps)$ become constants and we find $X_P(\varphi_{\sm}) = 0$,
i.e., the spin current $I_\sigma$ is insensitive to changes in $T_P$.
Interestingly, for the general case of energy-dependent densities of states our model predicts, quite generally,
a spin-dependent flow generated by nonzero $\Delta T_P$. To see this, let us
analyze the charge $I_c = I_{\up} + I_{\down}$ current,
\begin{multline}
I_c 
= \left[G(\varphi_{\up}) + G(\varphi_{\down})\right]\Delta V + \left[G(\varphi_{\up}) - G(\varphi_{\down})\right]\Delta V_s
\\
+ \left[L(\varphi_{\up}) + L(\varphi_{\down})\right]\frac{\Delta T}{T}
+ \left[X_{P}(\varphi_{\up}) + X_{P}(\varphi_{\down})\right]\frac{\Delta T_P}{T} \,,
\label{eq:Ic}
\end{multline}
and the spin $I_s = I_{\up} - I_{\down}$ current,
\begin{multline}
I_s 
= \left[G(\varphi_{\up}) - G(\varphi_{\down})\right]\Delta V + \left[G(\varphi_{\up}) + G(\varphi_{\down})\right]\Delta V_s
\\
+ \left[L(\varphi_{\up}) - L(\varphi_{\down})\right]\frac{\Delta T}{T}
+ \left[X_{P}(\varphi_{\up}) - X_{P}(\varphi_{\down})\right]\frac{\Delta T_P}{T} \,.
\label{eq:Is}
\end{multline}
In order to examine the possibility of generating pure spin current, i.e., $I_{c}=0$ and $I_{s}\ne0$, only via local heating ($\Delta T_P \ne 0$)
we set $\Delta T=0$ henceforth. Due to symmetry considerations,
we have $G(\varphi_{\up}) = G(\varphi_{\down})\equiv G(\varphi)$. In addition,
for symmetric couplings ($\Gamma_L=\Gamma_R$) we find
$X_{P}(\varphi_{\up}) =- X_{P}(\varphi_{\down})\equiv X_{P}(\varphi)$. Hence,
Eqs.~\eqref{eq:Ic} and \eqref{eq:Is} reduce to
\begin{align}
I_c &= 2G(\varphi)\Delta V \,,\\
I_s &= 2G(\varphi)\Delta V_s + 2X_{P}(\varphi)\frac{\Delta T_P}{T} \,.
\end{align}
This particularly simple result is central to our proposal. When no dc voltage is applied ($\Delta V = 0$),
the charge current $I_c$ vanishes and just a  pure spin current remains.
Furthermore, when the spin relaxation time in the reservoirs is very short,\cite{joh85}
we can neglect the term proportional to the spin bias $\Delta V_s$ and thus $I_s$ is uniquely determined by the temperature difference between
the phonon bath and the system. In general, our thermomagnetic device produces spin-polarized currents
from temperature differences only.

For realistic systems, lead couplings can be asymmetric and we then obtain nonvanishing charge currents.
However, this finite charge current can be cancelled by applying a compensating bias voltage across the system.
A similar compensation effect has been proposed for a two-terminal quantum dot system.\cite{rej12}
The difference is that we heat up the molecule using local phonon coupling and allow for (Rashba) electric fields
while Ref.~\onlinecite{rej12} uses Zeeman magnetic fields.
Thus, our proposed device is an {\em all-electrical} setup.
\begin{figure}[t]
\begin{center}
\includegraphics[width=9cm]{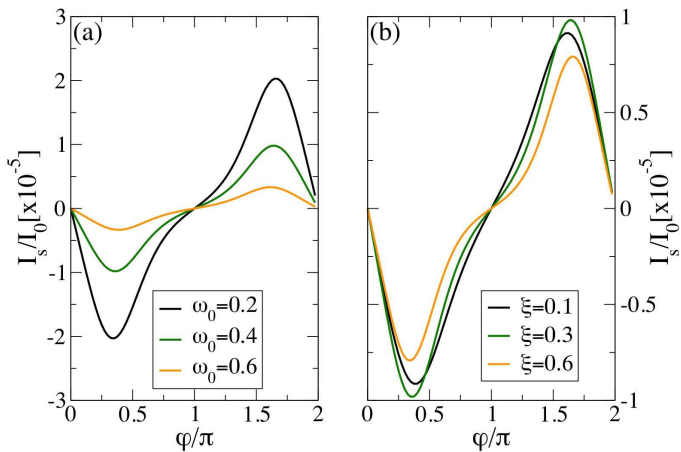}
\caption{Normalized spin current versus the Rashba phase $\varphi$ for (a) different phonon frequencies $\omega_0$
at fixed direct tunnel coupling $\xi=0.4$ and (b) different values of $\xi$ for $\omega_0=0.4$. Parameters: $\lambda=0.8$, $T=0.1$, $\Delta T_P=0.05$, $E_d=0.8$, and bandwidth $D=10$ with a semielliptic density of states $\rho(\varepsilon)=\sqrt{1-4(\varepsilon/D)^{2}}$. Here, energies are given in units of $\Gamma$ and the current unit is
$I_0=e\omega_0 \lambda^2/\Gamma^2$. }
\label{fig1}
\end{center} 
\end{figure}

\begin{figure}[t]
\begin{center}
\includegraphics[width=9cm]{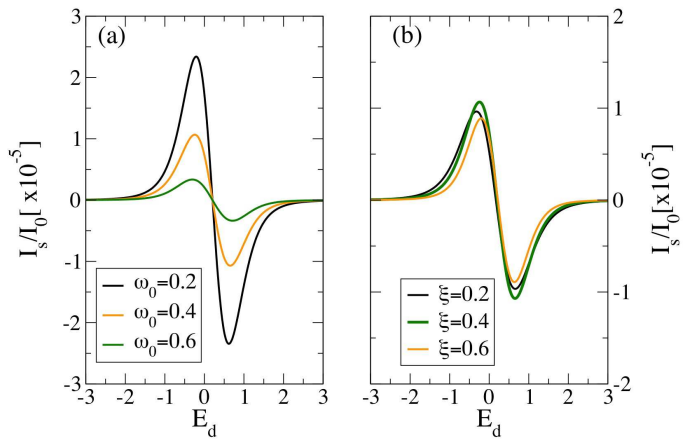}
\caption{Normalized spin current versus the energy level position $E_d$ for (a) different phonon frequency $\omega_0$
at fixed direct tunnel coupling $\xi=0.4$ and (b) different values of $\xi$ for $\omega_0=0.4$.
Parameters are taken as in Fig.~\ref{fig1} with a fixed Rashba phase $\varphi=0.35\pi$ near which the maximum spin current amplitude occurs.}
\label{fig2}
\end{center} 
\end{figure}

Figures~\ref{fig1} and \ref{fig2} display the spin currents as a function of the Rashba phase $\varphi$ and the molecular energy level $E_d$ for various values of the phonon frequency $\omega_0$ and direct tunneling strength $\xi$. Figure \ref{fig1} shows oscillations of the spin current when $\varphi$ changes from $0$ to $2\pi$. At fixed temperature, the number of phonons decreases when $\omega_0$ grows, yielding lower spin currents as shown in Fig.~\ref{fig1}(a). In Fig.~\ref{fig1}(b), we observe a weak dependence of the spin current amplitude on $\xi$. This is a nice property---the generated spin polarization is robust against unintentional variations
of the background transmission. Figure \ref{fig2} illustrates the gate dependence of $I_s$ at a fixed Rashba phase. We also find that the current amplitude enhances as $\omega_0$ decreases. Spin currents are maximized for specific positions of the molecular level: $E_d= -0.25\Gamma$, and $E_d=0.75\Gamma$. Moreover, the level positions at which $I_s$ is maximal depends weakly on $\omega_0$.
Finally, Fig.~\ref{fig2}(b)
presents small variation of the spin current when $\xi$ varies substantially, in agreement with the robustness discussed above.

In prototypical molecular transistors, vibrational frequencies are found to be of the order of THz for $C_{60}$\cite{park00}
and carbon nanotubes.\cite{leturq09} These systems are most suitable to test our predictions
due to their intrinsic spin-orbit coupling.\cite{minot00} We estimate from Figs.~2 and~3 that the achievable spin currents at equilibrium conditions are of the order of $I_s\sim 1$~pA for $\Gamma=\lambda=1$~meV, $\omega_0=1$~THz, $T=1$~K and $\Delta T_P=0.5$~K.
This current range is small but can be detected within present techniques.\cite{jav03}

In summary, we have proposed a spin current generator by locally heating a molecule
embedded in a spin-orbit interferometer. Our device works even in the absence
of voltage and temperature bias applied to the electronic terminals.
We have found that optimal spin polarization can be determined by adjusting the Rashba spin-orbit coupling
and the energy level of the molecule.
Our results are relevant in view of recent developments in the field of spin caloritronics.
Analogous setups could be envisaged where spin is replaced with orbital (pseudospin)
degrees of freedom.\cite{lim13}

The authors acknowledge the support from MECD
under Grants No. FIS2011-2352 and CSD2007-00042 (CPAN),
and the National Research Foundation (NRF) grant funded by the Korea government (MEST) (Grant No. 2011-0030790).

\end{document}